\documentclass[onecolumn,draftclsnofoot,12pt]{IEEEtran}

\usepackage{subfigure, amsmath}
\usepackage{setspace}
\usepackage{fullpage}
\usepackage{fancyhdr}
\usepackage{fullpage}
\usepackage{caption} 
\usepackage{hyperref}
\usepackage{multirow}
\usepackage{comment}
\usepackage{wrapfig}
\usepackage{amssymb}
\usepackage{amsmath}
\usepackage{amsfonts}
\usepackage{hyperref}
\usepackage{gensymb}
\usepackage{xcolor}
\usepackage[british]{babel}
\usepackage{hhline}
\usepackage{multirow}
\usepackage{adjustbox}
\usepackage{changepage}
\usepackage{graphicx}
\usepackage{romannum}
\usepackage{dcolumn}
\usepackage{bm}
\usepackage{nccmath}
\usepackage{commath}
\usepackage{braket}
\usepackage{booktabs}

\begin{document}
\pagenumbering{arabic}
\nocite{*}
\title{\Large \textbf{Capturing Protein Free Energy Landscape using Efficient Quantum Encoding}}
\author{\IEEEauthorblockN{\normalsize Ashwini Kannan$^{1}$, Jaya Vasavi Pamidimukkala$^{1}$,
Avinash Dakshinamoorthy$^{1}$, 
Soham Bopardikar$^{2}$,
Kalyan Dasgupta$^{3}$$^{*}$, and Sanjib Senapati$^{1}$$^{*}$}\\
\normalsize $^{1}$  Dept. of Biotechnology, Indian Institute of Technology Madras, Chennai, India\\
$^{2}$ Dept. of Electrical and Computer Engineering, Northeastern University, USA \\
$^{3}$IBM Research, Bangalore, India
\thanks{$^{*}$Corresponding author : Kalyan Dasgupta (email: kalyand1@in.ibm.com) ; Sanjib Senapati (email: sanjibs@iitm.ac.in) }}
\maketitle
\vspace{-1.5cm}
\begin{abstract}
Protein folding is one of the age-old biological problems that refers to the mechanism of understanding and predicting how a protein’s linear sequence of amino acids folds into its specific three-dimensional structure.This structure is critical, as a protein’s functionality is inherently linked to its final folded form. Misfolding can lead to severe diseases such as Alzheimer’s and cystic fibrosis, highlighting the biological and clinical importance of understanding protein folding mechanisms. This work presents a novel turn-based encoding optimization algorithm for predicting the folded structures of peptides and small proteins. Our approach builds upon our previous research, where our objective function focused on hydrophobic collapse, a fundamental phenomenon underlying the protein folding process. 
In this work, we extend that framework by not only incorporating hydrophobic interactions but also including all non-bonded interactions, such as van der Waals and electrostatic forces between residues, modeled using the Miyazawa–Jernigan (MJ) potential. We constructed a Hamiltonian from the defined objective function that encodes the folding process on a three-dimensional face-centered cubic (FCC) lattice, offering superior packing efficiency and a realistic representation of protein conformations. This Hamiltonian is then solved using classical and quantum solvers to explore the vast conformational space of proteins. To identify the lowest-energy folded configurations, we utilize the Variational Quantum Eigensolver (VQE), a hybrid quantum-classical algorithm, implemented on IBM’s 133-qubit hardware. The predicted structures are validated against experimental data using root-mean-square deviation (RMSD) as a metric and compared against classical simulated annealing and molecular dynamics simulation results. Our findings highlight the promise of hybrid classical-quantum approaches in advancing protein folding predictions, particularly for sequences with low homology.
\end{abstract}
\section{\textbf{INTRODUCTION}}
Proteins are biological macromolecules that perform a wide range of functions, including acting as receptors (ACE2), transporters (glucose transporters), and hormones (insulin), and a typical human cell contains approximately 20,000 unique proteins \cite{protein_human}. Proteins are composed of amino acids, and the way these amino acids fold into a specific three-dimensional structure determines the protein’s function, known as the protein folding process. Proper folding is essential to study, as its structure determines its stability and functionality. Misfolded proteins are associated with various diseases, including Alzheimer’s, Parkinson’s, sickle cell anemia, and cystic fibrosis. Understanding protein folding can help to elucidate disease mechanisms and develop therapeutic interventions for protein misfolding-related disorders. To date, 254 million protein sequences have been deposited in UniProt \cite{UniProt}; yet only around 230 thousand protein structures have been experimentally determined \cite{PDB} using techniques like X-ray crystallography, NMR, cryo-electron microscopy, etc. Given the complexity, cost, and time requirements of these experimental methods, computational approaches have emerged as indispensable tools for predicting protein structures, providing insights into folding pathways and dynamics at atomic resolution.

Protein structure prediction can be broadly classified into 2 categories: knowledge-based methods, where the structure of an unknown sequence is predicted based on the structures and sequences of existing proteins (e.g., AlphaFold \cite{alphafold}, RosettaFold \cite{rosetta}); and physics-based methods, where no homologous protein structures are available for effective structure prediction. While the first method is well-established and has achieved considerable success, our focus here is on the second approach. In physics-based methods, protein structure prediction relies solely on fundamental physical principles such as attaining the energy minima state. One common implementation of this approach is through molecular dynamics (MD) simulations \cite{MD}, which model the protein’s folding process over time. This approach begins with a random structure, and through energy minimization and equilibration in a simulated biological environment, the system converges toward the most stable, minimum energy conformation over time. 
Although advanced MD simulation techniques have shown promising results in protein structure prediction, the entire process is computationally complex and time-consuming, as highlighted by Leventhal’s Paradox \cite{Levinthal}. 

The enormous complexity of modeling protein folding can be reduced in two ways: (1) by discretizing the conformational space into a lattice framework, and (2) by coarse-graining the protein’s amino acid units. Several different lattice models have already been explored previously for studying protein folding that includes simple cubic (SC), body centered cubic (BCC), Face centered cubic (FCC) and tetrahedral lattices. Among all these, FCC has been shown to be best lattice for protein folding owing to its similarity to the intrinsic geometry of protein structures \cite{FCC}. The $C_\alpha-C_\alpha$ virtual bond angle in  proteins can take values of $90\degree$ in alpha-helix and $120\degree$ in beta-strands. Both these angles can be readily accommodated in FCC lattice. FCC lattice is also known to have the highest packing density among all the lattices which provides densest possible folded protein structures.

The second approach involves coarse-graining the protein structure into computationally tractable units suitable for quantum simulations. All proteins are polymers made up of 20 different amino acids each with its unique structure. Each amino acid can be coarse grained into single unit or beads and protein can be considered as linear sequences of connected beads. The interaction between these 20 amino acid beads can be modeled using knowledge based contact potentials derived from statistical sampling of available protein structures. Miyazawa-Jernigan is one such potential that is widely used for protein folding studies \cite{MJpot}. 

Since the reduced lattice model still has poor scalability when solved classically, researchers have started to explore the potential of quantum computing to provide more efficient and faster sampling, addressing the limitations of traditional methods.
Several attempts have been made to efficiently encode lattice structures into quantum computational basis states. \cite{resAnalysis_Linn} provides a details different encoding schemes and analysis of the hardware requirements for the protein lattice models. The encoding of the coordinate space to the computational basis states as proposed in \cite{Guzik_protein} is more expensive in terms of the required qubits, compared to turn-based encoding approaches proposed in many following papers \cite{Babbush_2014}, \cite{Anton_2021}. The cubic lattice was initially considered in many papers due to its simplicity \cite{Babbush_2014}, \cite{Babej_2018}. Tetrahedral lattices were considered in \cite{Anton_2021} for efficient resource (qubit) usage. These encoding approaches had limitations in their flexibility in generating larger number of turn angles or degrees of freedom. In \cite{doi:10.1021/acs.jctc.4c00848}, a turn-based encoding approach on a cubic lattice with high degrees of freedom was proposed. In \cite{dwaveQubo_Irback}, the authors have proposed a QUBO approach on a $4\times4\times3$ lattice using quantum annealing. In \cite{Omar_paper_FCC}, a turn-based encoding on the FCC lattice with higher degrees of freedom has also been proposed.  



In this paper, we present a novel turn-based protein 3D structure prediction model built on a Face-Centered Cubic (FCC) lattice that can be solved on a quantum computer. In order to decrease the computational and space complexity, we coarse-grained each amino acid into a single bead and explored multiple possible protein conformations, wherein each bead can occupy any of the available positions on the FCC lattice. The bead movement is governed by a potential function describing interactions between pairs of beads, along with few additional constraints that are crucial for ensuring realistic folding dynamics, are discussed elaborately in Section \ref{methodology}.

This paper is organized as follows: Section \ref{methodology} describes our methodology and the steps required to encode a classical optimization problem for quantum hardware. Section \ref{results} discusses the results from both classical and quantum algorithms run on IBM quantum simulators and hardware. Finally, in Section \ref{conclusions}, we present the essential conclusions from this study.
\section{\textbf{Methodology} \label{methodology}}

\begin{figure}[h]
\centering
\includegraphics[trim={0cm 10cm 0cm 8cm},clip, width=\linewidth]{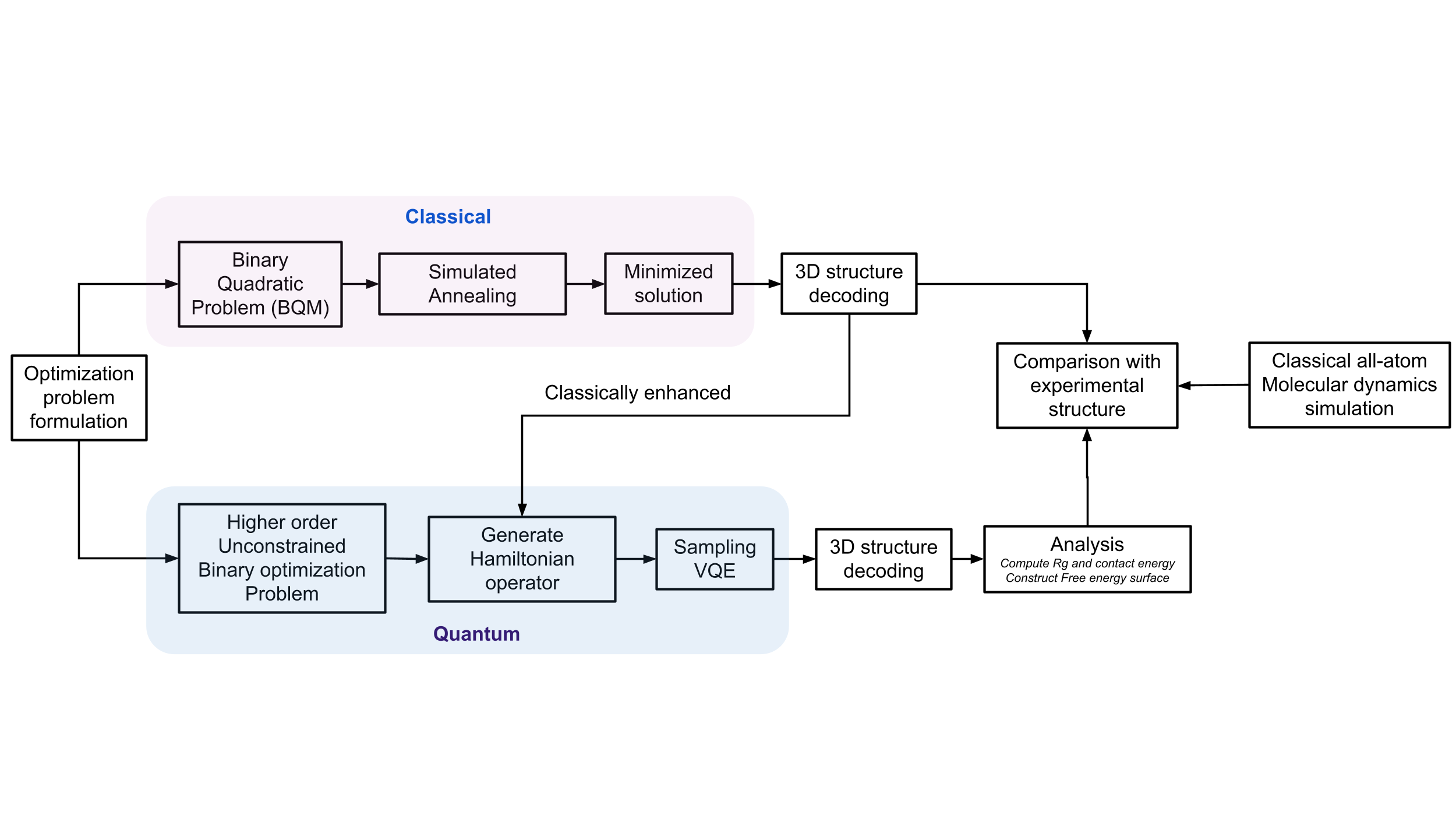}
\caption{The workflow for predicting the structure of protein by solving the optimization problem using Variational Quantum Eigensolver.}\label{fig:overview}
\end{figure}

In this work, we extend our previous study \cite{doi:10.1021/acs.jctc.4c00848}, which captures the hydrophobic collapse phenomenon. We formulate an optimization problem that minimizes the distance between amino acids according to specific interactions. In the case of the hydrophobic collapse problem, we had represented each amino acid as hydrophobic or hydrophilic beads, and the objective was to minimize the distances between hydrophobic beads. Here, we go beyond the binary representation, considering all 20 amino acids while still coarse-graining an amino acid as a bead centered at $C_\alpha$ atom. Figure \ref{fig:overview} captures the overall workflow adopted in this study to solve the optimization problem. Each component of the workflow is discussed below in detail.
\subsection{Optimization problem formulation \label{formulation}}
\subsubsection{Encoding on to a FCC lattice \label{fccEncoding}}
\begin{figure}[h]
\centering
\includegraphics[width=4.5in]{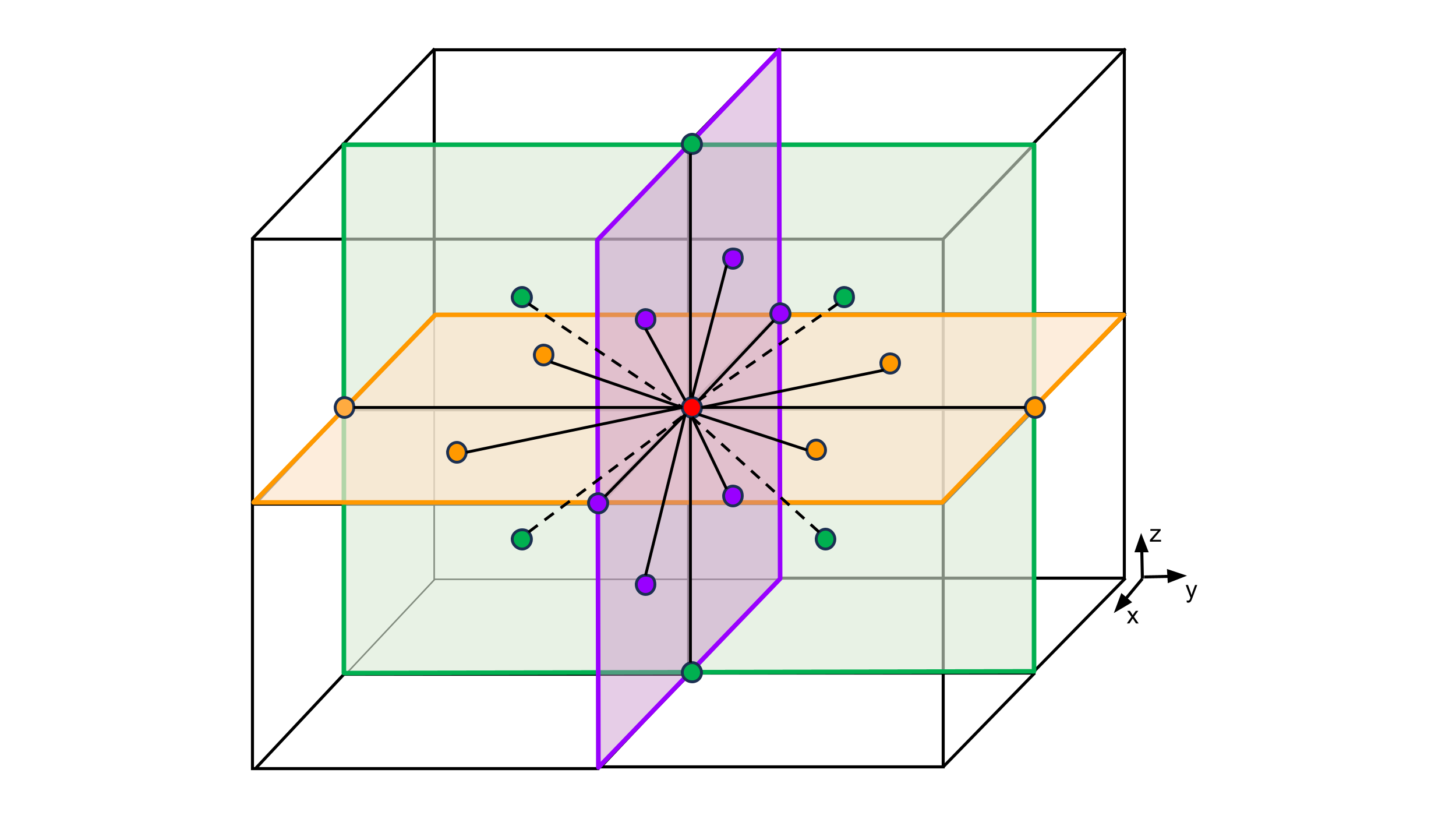}
\caption{An illustration of turn encoding in FCC lattice. Starting with the red bead at the centre, the 18 possible positions in the first turn have been split into 3 planes, i.e., 6 beads parallel to x-y plane (orange beads), 6 beads parallel to y-z plane (green beads), and 6 beads parallel to z-x plane (purple beads)}\label{fig:fcc}
\end{figure}

The folding problem is cast on a three-dimensional face-centered cubic (FCC) lattice, where we consider the first and second nearest neighbors, resulting in a total of 18 degrees of freedom (Figure \ref{fig:fcc}). A novel turn-based technique was introduced in \cite{fccEncoding} to encode the lattice structures in the computational basis states of qubits. However, due to the dependence on higher-order terms and the large number of terms in \cite{fccEncoding}, which in turn increases the number of Pauli strings in the Hamiltonian, we modified the encoding, as described below,
to reduce the overall complexity.

The 18 first and second neighbors are divided plane-wise, such as 6 neighbors in the x-y plane, 6 in the y-z plane, and 6 in the z-x plane. If we consider any bead $i$, say the red-colored bead in Figure \ref{fig:fcc} with coordinates $(x, y, z)$, the bead $i + 1$ would then have coordinates $(x + \Delta x, y + \Delta y, z + \Delta z)$. Depending on the plane chosen, it can take any one of the following:
\begin{flalign}
\begin{aligned}
\textrm{Parallel to the $y-z$ plane:}&~~ x_{i+1} = x_i, ~~~~~~~~~~~y_{i+1} = y_i + \Delta a_k,~~ z_{i+1} = z_i + \Delta b_k \\
\textrm{Parallel to the $z-x$ plane:}&~~ x_{i+1} = x_i + \Delta b_k,~~ y_{i+1} = y_i,~~~~~~~~~~~ z_{i+1} = z_i + \Delta a_k \\
\textrm{Parallel to the $x-y$ plane:}&~~ x_{i+1} = x_i + \Delta a_k,~~ y_{i+1} = y_i  + \Delta b_k,~~ z_{i+1} = z_i 
\end{aligned} \label{coord_update}
\end{flalign}
Since the first and second nearest neighbors are of distance 0.5 and 1 units from the $i^{th}$ bead respectively, the values of $\Delta a$ and $\Delta b$ can be,
$$\Delta a = \begin{pmatrix}
0.5\\
0.5 \\
-0.5\\
-0.5\\
0\\
0\\
0\\
0\\
\end{pmatrix} \Delta b = \begin{pmatrix}
0.5\\
-0.5\\
0.5\\
-0.5\\
1\\
0\\
0\\
-1\\
\end{pmatrix} $$
Overall we have 3 possible planes and 6 possible turns in each plane. Hence, for a single turn, we require at the least 2 qubits to choose a plane and 3 qubits to choose a turn in a plane; thus 5 qubits in total. For $N$ bead system, we require $5(N-1)$ qubits. Two qubits $(q_4,q_5)$ are used to choose one of the 3 planes; and 3 qubits $(q_1,q_2,q_3)$ are used to encode the turns.

\begin{figure}[h]
\centering
\includegraphics[width=3.5in]{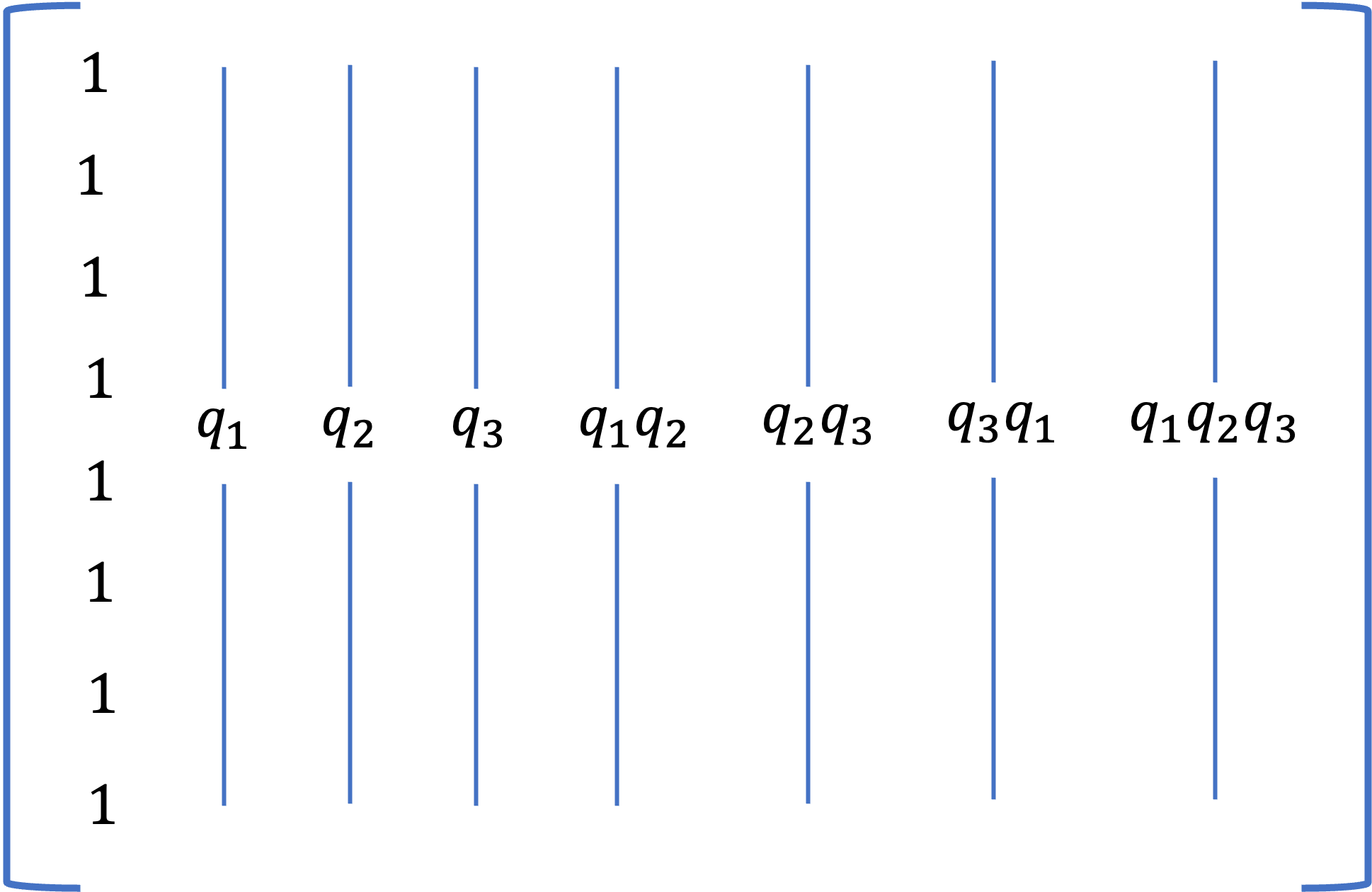}
\caption{Basis matrix structure for 3-qubit system}\label{fig:3qubitBasis}
\end{figure}

\textbf{Turn Encoding}: With 3 qubits $(q_1, q_2, q_3)$, choosing a turn from 8 possible states: \begin{equation}
    (q1,q2,q3) \in \{(0,0,0), (0,0,1), (0,1,0), (0,1,1), (1,0,0), (1,0,1), (1,1,0), (1,1,1)\}
    \label{3qubit}
\end{equation}
there are ${3\choose 1}$ ways to choose 1 qubit $(q_1, q_2, q_3)$, ${3\choose 2}$ ways to choose 2 qubits $(q_1q_2,q_2q_3,q_3q_1)$, and ${3\choose 3}$ ways to choose 3 qubits $(q_1q_2q_3)$. A basis matrix would then take a form as shown in Figure \ref{fig:3qubitBasis}.

Substituting \ref{3qubit}, the basis matrix $B_3$ can be written as 
\begin{equation}
    \begin{aligned}
        B_3 =
\begin{pmatrix}
\hphantom{0}1\hphantom{0} & \hphantom{0}0\hphantom{0} & \hphantom{0}0\hphantom{0} & \hphantom{0}0\hphantom{0} & \hphantom{0}0\hphantom{0} & \hphantom{0}0\hphantom{0} & \hphantom{0}0\hphantom{0} & \hphantom{0}0\hphantom{0}\\
1 & 0 & 0 & 1 & 0 & 0 & 0 & 0\\
1 & 0 & 1 & 0 & 0 & 0 & 0 & 0\\
1 & 0 & 1 & 1 & 0 & 1 & 0 & 0\\
1 & 1 & 0 & 0 & 0 & 0 & 0 & 0\\
1 & 1 & 0 & 1 & 0 & 0 & 1 & 0\\
1 & 1 & 1 & 0 & 1 & 0 & 0 & 0\\
1 & 1 & 1 & 1 & 1 & 1 & 1 & 1\\
\end{pmatrix}
    \end{aligned}
\end{equation}
The vectors $\Delta a$ and $\Delta b$ can be represented by basis $B_3$ as 
\begin{equation}
    \begin{aligned}
        [\Delta a, \Delta b] = [c_{\Delta a}^\dagger B_3, c_{\Delta b}^\dagger B_3]
    \end{aligned}
\end{equation}
The coefficients are estimated from 
\begin{equation}
    \begin{aligned}
        [c_{\Delta a}, c_{\Delta b}] = [B_3^{-1} \Delta a, B_3^{-1} \Delta b]
    \end{aligned}
\end{equation}

The coefficients obtained are, 
\begin{equation}
    \begin{aligned}
        c_{\Delta a} &= (0.5, -0.5, 0, -1, 0, 0, 1, 0)\\
        c_{\Delta b} &= (0.5, -0.5, -1, 0, 0, 0, -1, 0)
    \end{aligned}
\end{equation}
And hence,

\begin{equation}
    \begin{aligned}
        \Delta a = 0.5 - 0.5q_1 - q_2 + q_1 q_2 \\
 \Delta b = 0.5 + 0.5 q_1 - q_3 - q_1q_2 \\
    \end{aligned}
\end{equation}

\textbf{Plane encoding}: To choose one of 3 planes, we need a minimum of 2 qubits $(q_4, q_5) \in \{00,01,10,11\}$. The qubit encoding is as follows:

\begin{itemize}
    \item Parallel to x-y plane: $(q_4, q_5) = (1,1) : q_4 q_5$ 
    \item Parallel to y-z plane: $(q_4, q_5) = (0,1) : (1-q_4)q_5$ 
    \item Parallel to z-x plane: $(q_4, q_5) = (1,0) : q_4(1 - q_5)$
\end{itemize}

Combining both the plane and turn encoding, we have

\begin{equation}
    \begin{aligned}
        x_{i+1} &= x_i + q_4 q_5\Delta a + q_4(1 - q_5)\Delta b \\
 y_{i+1} &= y_i + (1-q_4)q_5\Delta a+ q_4q_5\Delta b \\
z_{i+1} &= z_i + q_4(1- q_5)\Delta a + (1-q_4)q_5\Delta b 
    \end{aligned}
\end{equation}
Now the turn taken from $i^{th}$ residue to the $i+1^{th}$ residue, represented by $(x_t, y_t, z_t)$ can be obtained from 

\begin{equation}
    \begin{aligned}
        x_t &= x_{i+1} - x_i \\
        y_t &= y_{i+1} - y_i \\
        z_t &= z_{i+1} - z_i
    \end{aligned}
    \label{coordToturn}
\end{equation}

\subsubsection{Hamiltonian formulation \label{hamiltonianFormulation}}
The objective function is constructed to minimize the sum of all pairwise distances between non-adjacent amino acids depending on the Miazawa-Jernigan (MJ) potential. Furthermore, the constraints are incorporated into the objective to penalize  the discontinuity $(C_{1})$, overlap $(C_2)$, and diagonal crossing $(C_3)$ (crossing of bonded pairs) of the beads, as was used in our previous study \cite{doi:10.1021/acs.jctc.4c00848}. The consolidated objective function is expressed as follows:
\begin{eqnarray}
    \min \lambda_0Obj + \lambda_1 C_{1} - \lambda_2 C_2 - \lambda_3 C_3 \label{eqn:final_form}
\end{eqnarray}
where,
\begin{eqnarray}
Obj = \sum_{j=1}^{N-2} \sum_{k=j+2}^{N} w_{jk}\Bigg[ \left(\sum_{l=j}^{k-1} x_t^l\right)^2 + \left(\sum_{l=j}^{k-1} y_t^l\right)^2  + \left(\sum_{l=j}^{k-1} z_t^l\right)^2 \Bigg ] \label{obj-main}
\end{eqnarray}

\begin{multline}
    C_1 = \sum_{i=1}^{N-1} \Big[ 1 -  \left(x_t^i\right)^2 -  \left(y_t^i\right)^2 -  \left(z_t^i\right)^2 +  \left(x_t^i\right)^2 \left(y_t^i\right)^2 \\
    + \left(y_t^i\right)^2 \left(z_t^i\right)^2\ + \left(z_t^i\right)^2 \left(x_t^i\right)^2 - \left(x_t^i\right)^2 \left(y_t^i\right)^2\left(z_t^i\right)^2 \Big] \label{continuity_const} 
\end{multline}

\begin{equation}
     C_2 = \sum_{i=1}^{N-2} \sum_{j=i+2}^{N} \Bigg[ \alpha _{ij}\left(\sum_{l=i}^{j-1} x_t^l\right)^2 + \beta _{ij}\left(\sum_{l=i}^{j-1} y_t^l\right)^2 
     + \gamma _{ij}\left(\sum_{l=i}^{j-1} z_t^l\right)^2 \Bigg] \label{overlap_const}
\end{equation}

\begin{equation}
    C_3 = \sum_{r=1}^{N-3} \sum_{k=r+2}^{N-1} \left[ \alpha _{rk}X_{rk}^2 + \beta _{rk}Y_{rk}^2 + \gamma _{rk} Z_{rk}^2\right]
\end{equation}
with N being the number of amino acid (or beads) in the protein sequence. The weights $w_{jk}$ in \ref{obj-main} is the pairwise MJ potential corresponding to $j^{th}$ and $k^{th}$ beads.  The coefficients $\lambda_0, ~\lambda_1, ~\lambda_2, ~\lambda_3 $ in the consolidated objective \ref{eqn:final_form} are the penalty factors for the objective and constraint functions. The following penalty factors are chosen: $\lambda_0 = 1$, $\lambda_1 = \abs{\frac{c_{obj}}{c_{continuity}}}$, $\lambda_2 = \abs{\frac{c_{obj}}{c_{overlap}}}$, and $\lambda_3 = 0.5 \times \lambda_2$, with $c_{obj} = \sum_{jk} w_{jk}$ and $c_{continuity}$, $c_{overlap}$ are the total number of possible terms in $C_1$ and $C_2$, respectively. The coefficients $(\alpha, \beta, \gamma)$ in the constraints $C_2$ and $C_3$ are randomly chosen from the normal distribution such that at least one of them is 1 and others are 0.
\subsection{Classical and quantum approaches \label{Approaches}}
The objective function is constructed using a modified PyQUBO library to accommodate higher-order terms. Classically, the optimization problem is solved using simulated annealing. First, the objective function is approximated to quadratic order by neglecting higher-order contributions. This quadratic form is then converted into a binary quadratic model (BQM) object, which is subsequently solved using PyQUBO’s built-in annealing samplers.

For the quantum approach, the Hamiltonian operator is constructed by transforming the binary objective into strings of the Pauli matrix $Z$ and the identity $I$. The objective is then minimized using the Variational Quantum Eigensolver (VQE), \ref{vqe} along with conditional value-at-risk (CVaR) \cite{Barkoutsos-cVar}.
\begin{equation}
    \min_{\theta} \braket{\psi_\theta|H|\psi_\theta}
    \label{vqe}
\end{equation}
The expectation value is optimized with respect to the ansatz parameter $\theta$ using the COBYLA optimizer. 

Due to the limitations in the current hardware, the experiments are carried out in a hybrid manner. Initial VQE iterations are run on a Matrix Product State (MPS) simulator until the convergence is reached. The experiments on the hardware are run for 50 iterations starting from the optimal parameter of the simulator run. The number of shots per iteration is set to 4000. Due to the large sampling space, for example a 10-bead sequence requires 45 qubits which makes exhaustive sampling impractical, all bitstrings generated from the hardware runs are retained for further analysis.
\subsection{3D structure Decoding \label{decoding}}
The bitstrings obtained from both classical simulated annealing and VQE are converted to 3D coordinates from \ref{coordToturn}. The resulting coordinates are then adjusted so that adjacent beads are separated by a uniform distance of $3.8 \AA$ (corresponding to $C_{\alpha}-C_{\alpha}$ distance), while preserving the original bond angles between successive beads. The root mean square deviation (RMSD) of the $C_{\alpha}$ atoms computed with respect to the experimental structure using 
\begin{equation}
    RMSD = \sqrt{\frac{1}{N} \sum_{i=1}^N \norm{\textbf{r}_i^m - \textbf{r}_i^{exp}}^2}
    \label{rmsd}
\end{equation}

In the case of VQE, the bitstrings from hardware run are first examined for structural violations such as overlaps and diagonal crossings, and only those containing at most one violation are selected for correction. These violations are then corrected by relocating the overlapping bead to any of its nearest valid neighbors on the lattice. For these filtered bitstrings, the contact energy and the radius of gyration ($R_g$) are calculated. The radius of gyration is computed as
\begin{equation}
    R_g = \sqrt{\frac{1}{N} \sum_{i=1}^N \norm{\textbf{r}_i - \textbf{r}_{cm}}^2}
    \label{rg}
\end{equation}
where $\textbf{r}_{cm} = \frac{1}{N} \sum_i \textbf{r}_i$ denotes the center of mass of the structure. The contact energy computed for non-adjacent bead pairs separated by less than $8\AA$ \cite{dist_cutoff}, 
\begin{equation}
    E_{\mathrm{contact}} = \sum_{j=1}^{N-2} \sum_{k=j+2}^N w_{jk} \hspace{0.2cm} \Theta(8 - \norm{\textbf{r}_j - \textbf{r}_k}),
    \label{contact energy}
\end{equation}
where $\Theta$ is the Heaviside step function and $w_{jk}$ is the MJ potential. Finally, the RMSD of the $C_{\alpha}$ atoms using \ref{rmsd} is calculated with $\textbf{r}^m$ being the corrrdinates corresponding to the minimum free energy bin on the contact energy–$R_g$ free energy surface with respect to the experimental structure.

For classical simulated annealing, $\textbf{r}^m$  in \ref{rmsd} corresponds to the 3D coordinates of minimized output bitstring.

\subsection{Classical Molecular Dynamic Simulations protocol}

To compare the quantum results with current standard non-homologous peptide structure prediction methods, we performed all-atom molecular dynamics (MD) simulations to explore the folding mechanisms and identify the lowest energy native conformation of selected peptides. The linear peptide structures were initially generated using Chimera \cite{chimera}, after which the systems were solvated with water and neutralized with ions to mimic a physiological environment at 0.15 M ionic concentration. This was followed by energy minimization using the steepest descent algorithm to remove any steric clashes or unfavorable contacts. Subsequently, the systems were gradually heated to 310 K using the V-rescale thermostat \cite{v_rescale}, and isothermal-isobaric equilibration was carried out at 1 bar pressure using the Parrinello–Rahman barostat \cite{parrinello1981polymorphic}. Following equilibration, we proceeded to the production runs using the AMBER-99SB-ILDN force field \cite{amber_ff} within the GROMACS 2024.3 package \cite{gromacs_2024}. A 300 ns production MD simulation was conducted in the NPT ensemble to capture the conformational dynamics of the peptides in solution over time and to identify their lowest energy states. Electrostatic interactions were treated using the Particle Mesh Ewald (PME) method with a 1.0 nm cutoff, while van der Waals interactions were truncated beyond the same cutoff. The LINCS algorithm was applied to constrain all bonds involving hydrogen atoms, allowing the use of a 2 fs integration time step.

After completion of the production runs, the simulation trajectories were analyzed to identify the most representative peptide conformations. Two complementary approaches were employed: clustering-based analysis and potential energy–radius of gyration (Rg) mapping.

For the clustering-based method, equilibrated portions of the trajectories were extracted and subjected to RMSD-based clustering using a cutoff of 1.5~\AA. All frames within this threshold were grouped into the same cluster after aligning the peptide backbone atoms. The centroid structure of the most populated cluster was considered the representative, most probable, and energetically favorable conformation of the peptide. In the second approach, a two-dimensional potential energy vs radius of gyration (Rg) map was generated over the entire trajectory to visualize the conformational free-energy landscape. The average structure corresponding to the high-density (low-energy) region of the map, as shown in Figure~\ref{fig:FES}, was extracted and compared with experimental structures.

Once the predicted structures were obtained from the MD simulations, they were compared against the corresponding experimental structures, which served as controls. The structural similarity was quantified by calculating the root mean square deviation (RMSD) between the predicted and experimental conformations, as defined in Equation~\ref{rmsd}, after aligning the C\textsubscript{$\alpha$} atoms of both structures.

\section{\textbf{Results} \label{results}}

To validate our encoding and optimization framework, we selected a set of peptides with known experimental structures determined by X-ray crystallography or solution NMR. The peptides chosen span lengths of 6 to 10 amino acids, and additional longer peptides ranging from 11 to 20 amino acids were also included. The following subsections present and discuss the results obtained using the encoding scheme described in the Methodology section.

\subsection{Simulator-Hardware implementation}

\begin{figure*}[h]
\centering 
\includegraphics[trim={0cm 22cm 12cm 0cm}, clip, width=\textwidth]{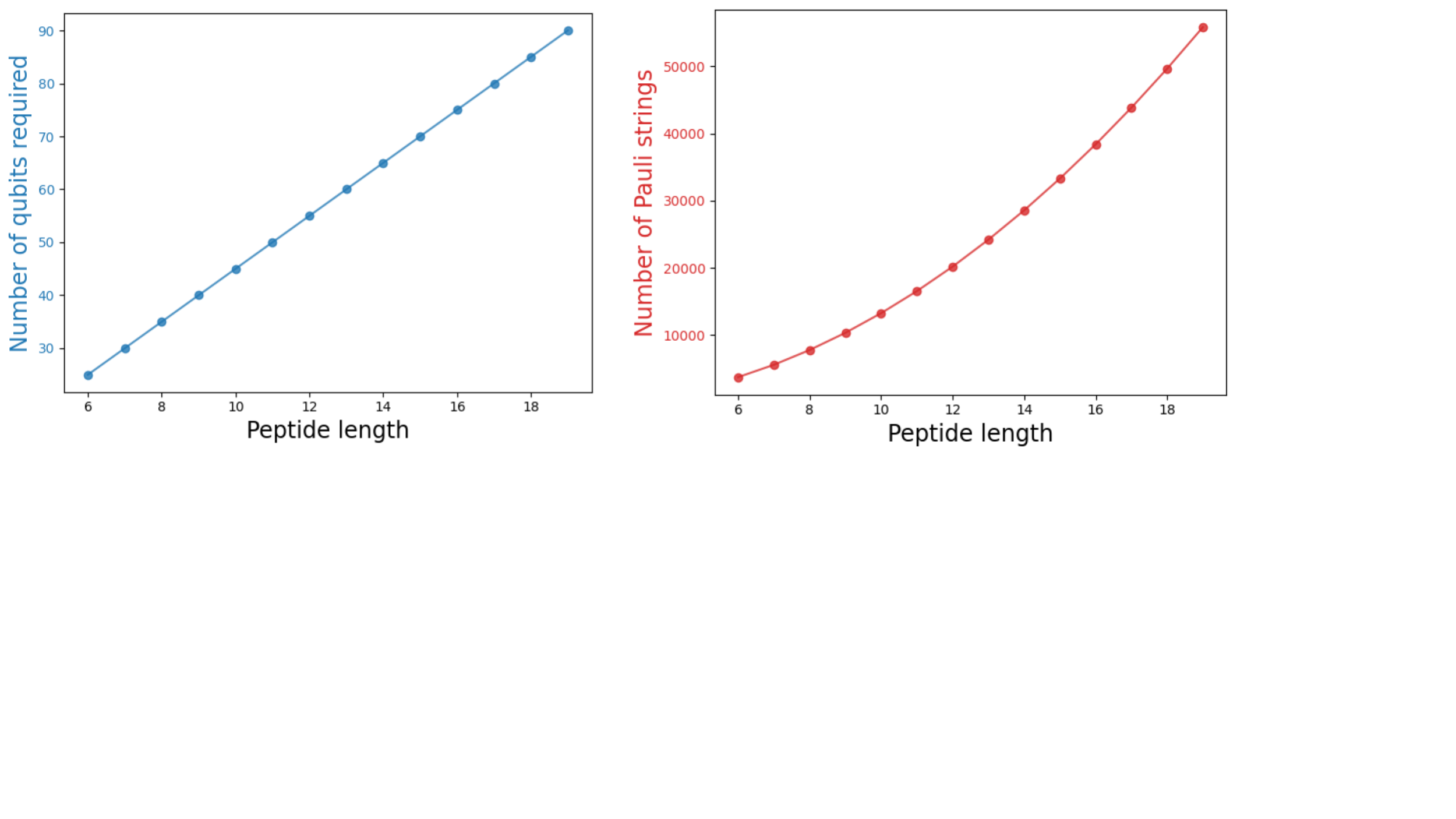} 
\caption{ Scaling of our encoding scheme: Qubit requirements (blue) and number of Pauli string in the Hamiltonian (red) scale up as  $\mathcal{O}(N)$ and $\mathcal{O}(N^2)$ respectively.}
\label{fig:scale}
\end{figure*}

According to our encoding method mentioned in Section \ref{fccEncoding}, each k-local term expands to $2^{4k}$ Pauli strings. The objective function, along with the overlap and diagonal crossing constraints, comprises two-local terms and exhibits a computational scaling of order $\mathcal{O}(N^2)$. The continuity constraint involves six-local terms and scales as $\mathcal{O}(N)$. The overall quadratic scaling of the number of Pauli strings and the linear scaling of the number of qubits with respect to the peptide length is shown in Figure \ref{fig:scale}.

\begin{figure*}[h]
\centering 
\includegraphics[trim={0cm 25cm 55.5cm 3cm},clip,width=0.6\textwidth]{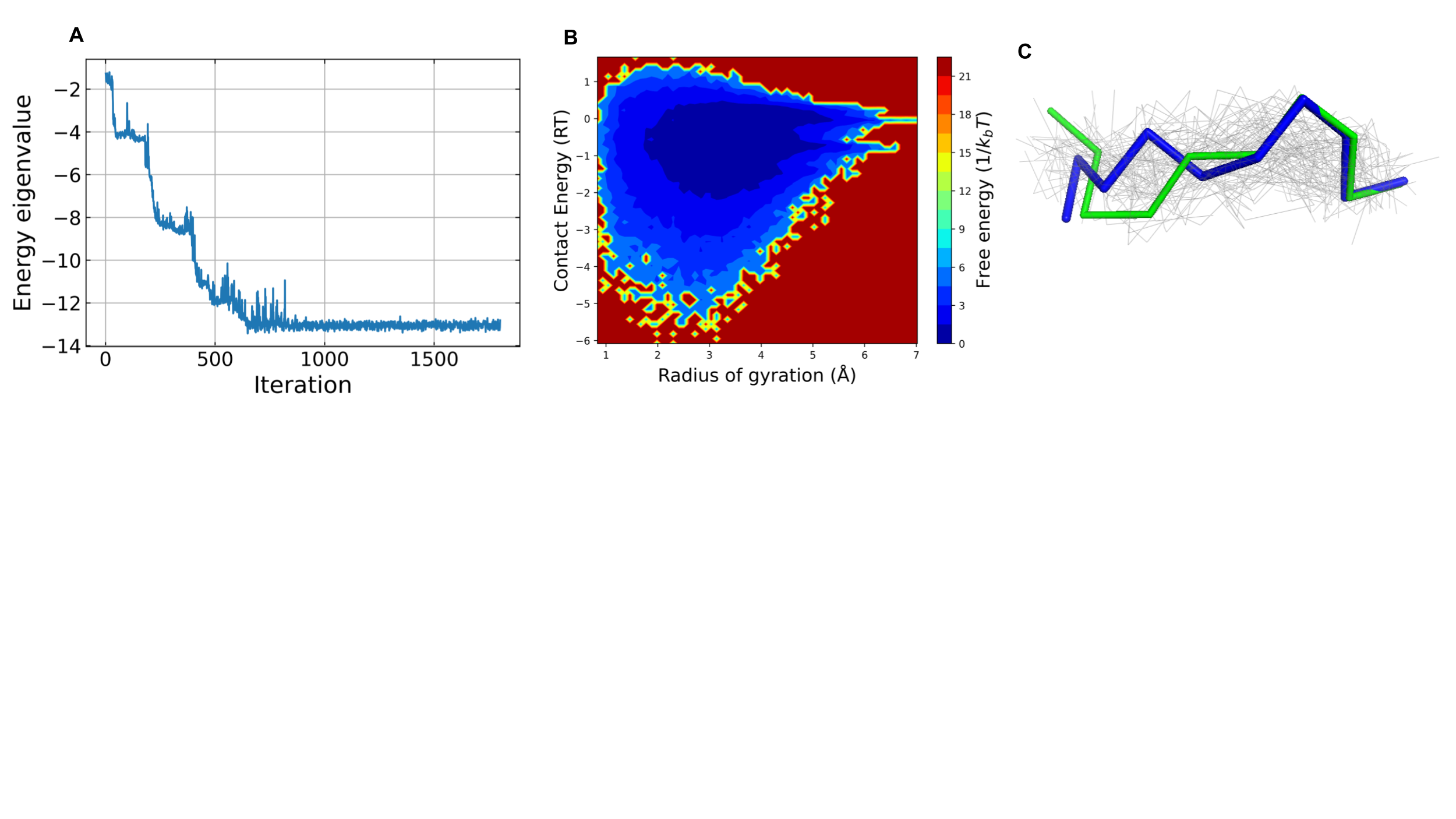} 
\caption{Convergence plot from the VQE iterations for Angiotensin (DRVYIHPFHL). These are run on the simulator and the final hardware runs are initiated with these converged parameters.}
\label{fig:iters}
\end{figure*}

\begin{figure*}[h]
\centering 
\includegraphics[trim={0cm 1cm 0cm 1cm},clip,width=\textwidth]{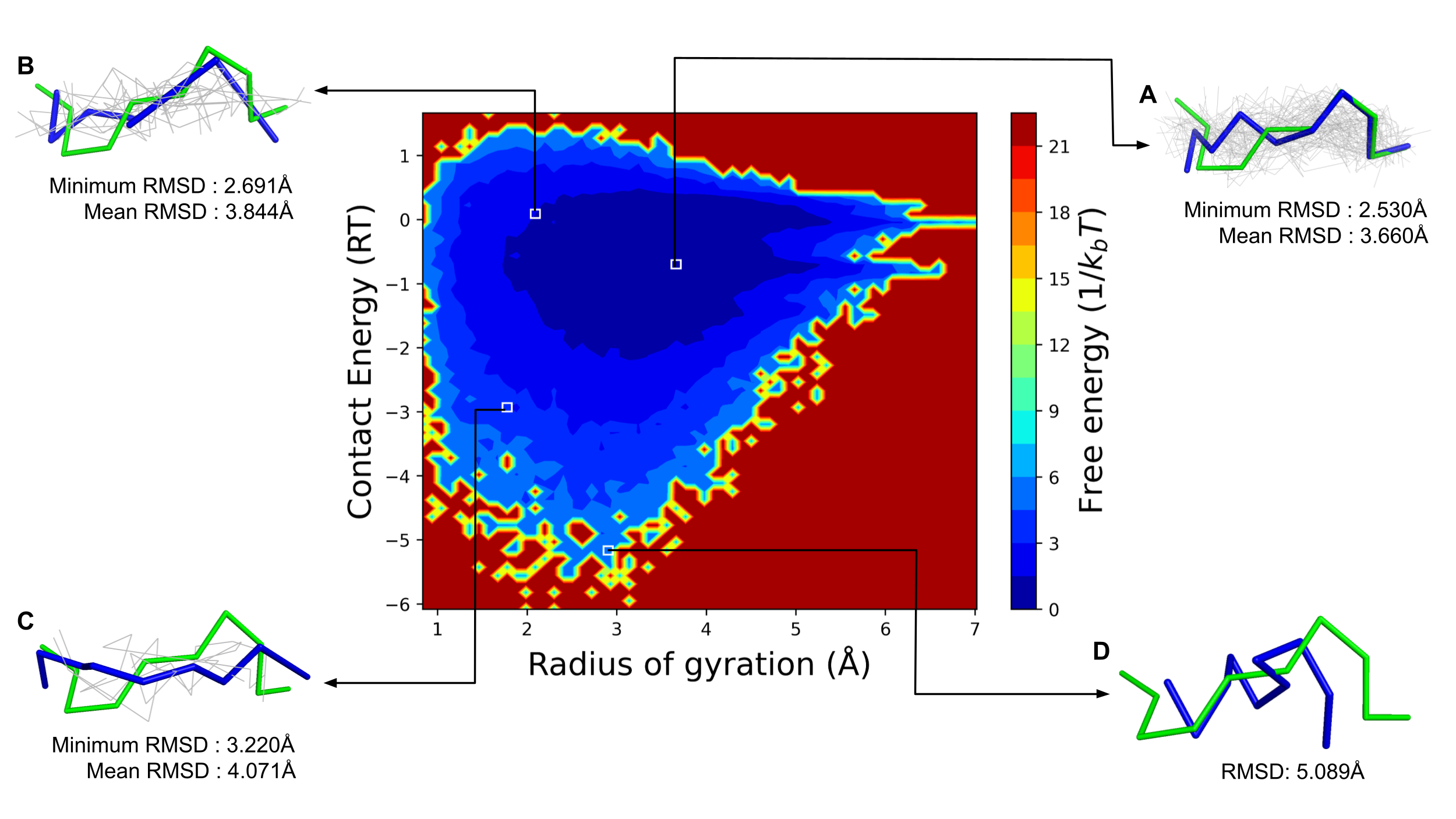} 
\caption{ Free energy landscape between contact energy $E_\text{contact}$ and the radius of gyration $R_g$ obtained from the hardware run for Angiotensin. The structures (A)-(D) corresponds to the minimum free energy bin in each contour. Green indicates the experimental backbone, blue indicates the minimum RMSD structure within respective bin, and gray indicates the remaining structures in that bin.}
\label{fig:FES}
\end{figure*}

\begin{figure*}[h]
\centering 
\includegraphics[width=\textwidth]{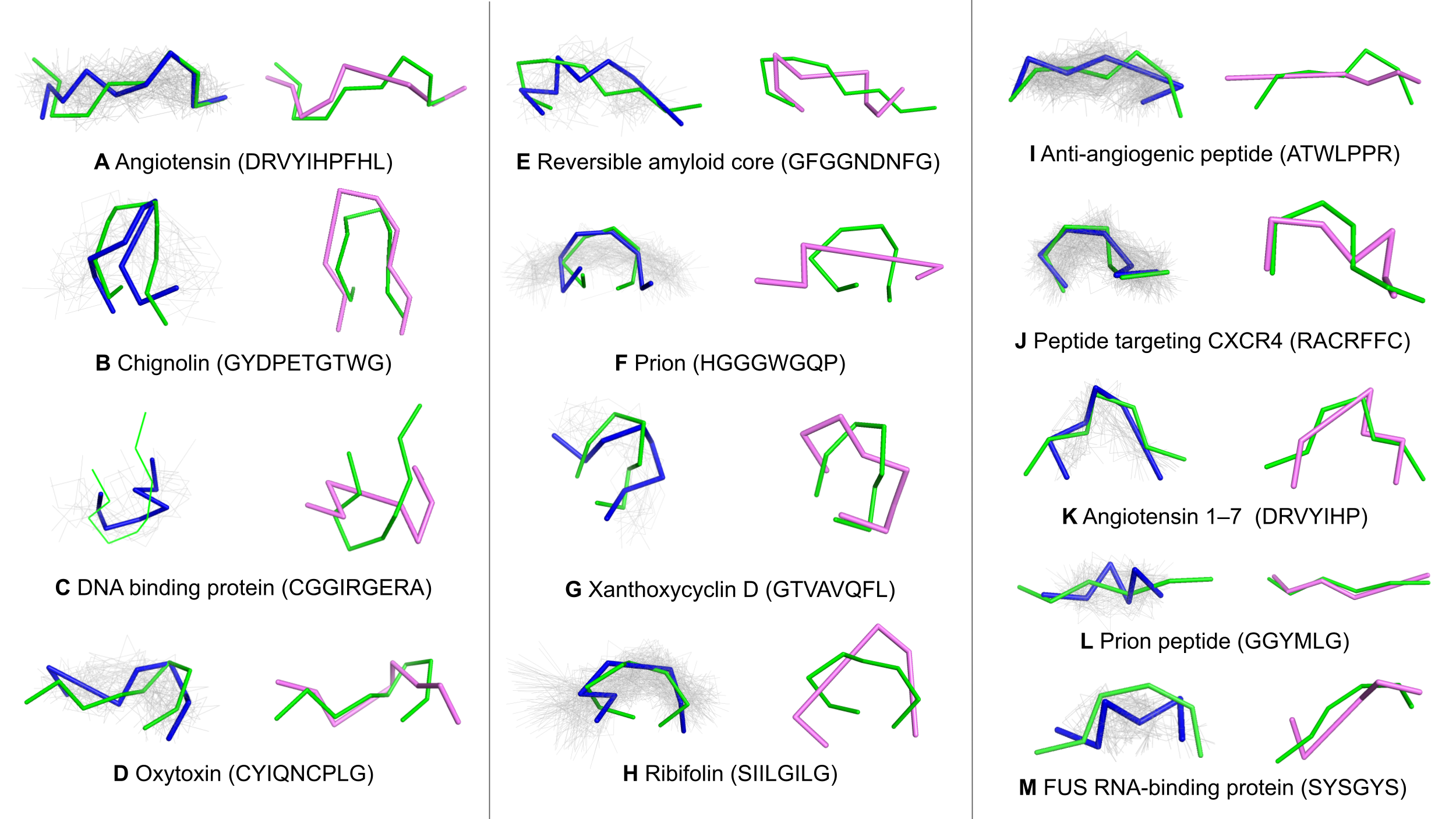} 
\caption{The predicted structures of 13 peptides: green indicates the experimental backbone, blue is the minimum RMSD structure within the minimum free energy bin. Gray lines are the remaining structures in that bin, and pink indicates the minimum structure obtained from classical simulated annealing.}
\label{fig:finalStructures}
\end{figure*}

Following the approach described in Section \ref{Approaches}, the VQE method was first executed on a simulator to reach a preliminary convergence, after which the final 50 iterations were performed on the hardware. Hardware experiments are conducted on the 133-qubit IBM Heron processor (\textit{ibm\_torino}). All the simulator and hardware implementation were carried out using Qiskit SDK \cite{qiskit}. We employed the 'Efficient SU(2)' ansatz, with the simulator run initialized using random parameters. We used M3 error correction \cite{mthree} to mitigate the sampling error in the VQE Sampler for the hardware run. Figure \ref{fig:iters} shows the convergence plot from the simulator run for Angiotensin. It can be seen that it took about 800 iterations for the energy eigenvalue to converge to the global minima. The 2D free energy landscape obtained from the hardware experiments is shown in Figure \ref{fig:FES}. The structures corresponding to the minimum bin for each contour are shown (A)-(D) subfigures. It can be seen that the average RMSD tends to decrease as the free energy increases. The representative backbone structures are extracted from the global minima in the free energy surface as these represent the dominant conformations sampled. The resulting structures corresponding to all 13 peptides are shown in Figure \ref{fig:finalStructures}.

\begin{table}[h]
    \centering
    \caption{Root Mean Square deviation (RMSD) comparison of structures obtained using IBM machines, classical simulated annealing (SA), and classical molecular dynamics (MD), each evaluated with respect to the experimental reference structure.}
    \label{tab:model_comparison}
    \renewcommand{\arraystretch}{1.2} 
    \begin{adjustbox}{width=\textwidth} 
    \begin{tabular}{lcccccc}
        \toprule
        \multirow{3}{*}{Peptide} & \multirow{3}{*}{PDB id  } & \multirow{3}{*}{Sequence} &  \multicolumn{4}{c}{RMSD ($\AA$)}\\
        \cmidrule(lr){4-7}
        &&&\multicolumn{2}{c}{IBM Hardware} &\multirow{2}{*}{\hspace{0.5cm}SA\hspace{0.5cm}} &\multirow{2}{*}{\hspace{0.5cm}MD\hspace{0.5cm}}\\
        \cmidrule(lr){4-5}
        &&&\hspace{0.2cm}Minimum\hspace{0.2cm} &\hspace{0.2cm}Mean\hspace{0.2cm}&  & \\
\midrule
 \multirow{1}{*}{Angiotensin} & \multirow{1}{*}{1N9U} & \multirow{1}{*}{DRVYIHPFHL  } & 2.530 & 3.660 & \textbf{1.964} & 5.11 \\
\multirow{1}{*}{Chignolin} & \multirow{1}{*}{1UAO} & \multirow{1}{*}{GYDPETGTWG  } & 2.775 & 3.908 & 2.190 & \textbf{0.79} \\
\multirow{1}{*}{DNA binding protein} & \multirow{1}{*}{1CS9} & \multirow{1}{*}{CGGIRGERA} & 2.223 & 3.467 & 3.924 & \textbf{1.826}   \\
\multirow{1}{*}{Oxytocin} & \multirow{1}{*}{2MGO} & \multirow{1}{*}{CYIQNCPLG} & \textbf{2.000} & 3.160 & 2.289 & 2.788 \\
\multirow{1}{*}{Reversible amyloid core} & \multirow{1}{*}{5ZGD} & \multirow{1}{*}{GFGGNDNFG} & \textbf{2.685} & 3.988 & 2.983 &  5.32 \\
\multirow{1}{*}{Prion} & \multirow{1}{*}{1OEH} & \multirow{1}{*}{HGGGWGQP} & \textbf{1.969} & 4.334 & 4.745 &  2.974\\
\multirow{1}{*}{Ribifolin} & \multirow{1}{*}{6DKZ} & \multirow{1}{*}{SIILGILG} & \textbf{2.142} & 4.190 & 2.765 & 2.658 \\
\multirow{1}{*}{Xanthoxycyclin D} & \multirow{1}{*}{6WPV} & \multirow{1}{*}{GTVAVQFL} & 3.060 & 3.579 & \textbf{2.395} & 3.632 \\
\multirow{1}{*}{Anti-angiogenic peptide} & \multirow{1}{*}{2JP5} & \multirow{1}{*}{ATWLPPR} & 2.075 & 2.839 & 2.203 & \textbf{1.616}  \\
\multirow{1}{*}{Peptide targeting CXCR4} & \multirow{1}{*}{5LFF} & \multirow{1}{*}{RACRFFC} & \textbf{1.220} & 2.630 & 2.871 & 2.29 \\
\multirow{1}{*}{Angiotensin 1--7} & \multirow{1}{*}{2JP8} & \multirow{1}{*}{DRVYIHP} & 2.043 & 2.759 & \textbf{1.790} & 2.36 \\
\multirow{1}{*}{Prion peptide} & \multirow{1}{*}{4TUT} & \multirow{1}{*}{GGYMLG} & 3.115 & 4.490 & \textbf{0.731} &  3.826 \\
\multirow{1}{*}{FUS RNA-binding protein} & \multirow{1}{*}{5XSG} & \multirow{1}{*}{SYSGYS} & \textbf{2.189} & 4.004 & 2.759 & 3.092 \\
        \bottomrule
    \end{tabular}
    \end{adjustbox}
    \label{RMSDEffSU2}
\end{table}

Table \ref{RMSDEffSU2} summarizes the RMSD values for 13 peptides, comparing the structures predicted using IBM machines with those obtained from classical simulated annealing and molecular dynamics simulations. Across all peptides studied, the minimum RMSD structures predicted using quantum computers consistently lie in the range $1.22\AA$ to $3.11\AA$ of the corresponding experimental structures. For shorter peptide sequences, classical simulated annealing achieves RMSD values that are comparable to those obtained with IBM, and in a few cases, such as the Prion peptide, even slightly better performance is observed.

\subsection{Custom ansatz}

To reduce the number of parameters and accelerate convergence, we explored alternative, non-standard ansatz. Figure 
illustrates a custom-modified version of the efficient SU(2) ansatz, designed specifically to accommodate our encoding scheme. While the minimum RMSD structures obtained using the standard efficient SU(2) ansatz showed good alignment with experimental structures, the other conformations within the minimum free energy bin deviated substantially from the known structure. Using the custom ansatz, we observed improvements in the minimum RMSD as well as a reduction in the mean RMSD across the bin.

\subsection{Classical MD simulations}

To evaluate the performance of our quantum approach and assess its consistency with established classical methods, we performed molecular dynamics (MD) simulations for all peptides listed in Table~\ref{tab:model_comparison}, starting from an extended linear structure. The production runs were carried out for 300–600~ns, depending on the stability of each system over time. The simulations were terminated once the structures reached equilibrium and were stable for a long time, after which several analyses were performed to obtain the native predicted structure of each peptide, such as clustering, RMSD vs gyration 2D plots, Free energy landscape, Potential energy vs gyration 2D plots, etc. The resulting structures were compared with their corresponding experimental conformations using the root mean square deviation (RMSD) as the evaluation metric.

Once equilibrium was achieved, stable frames from the trajectories were extracted, followed by clustering based on structural similarity. The most populated cluster centroid was considered the final predicted structure from the MD simulations. The RMSD values corresponding to these structures are presented in the MD column of Table~\ref{tab:model_comparison} (Results obtained from the peptide structures identified from the potential energy vs gyration 2D plots are presented in Supplementary Table [ref]). We observed that peptides such as Chignolin, Anti-angiogenic peptide, and DNA-binding protein exhibited close agreement with their experimental structures, showing RMSD values of 0.79~\AA, 1.616~\AA, and 1.826~\AA, respectively. In contrast, peptides such as Angiotensin and Reversible Amyloid Core displayed higher deviations with RMSD values of 5.11~\AA~and 5.32~\AA, respectively. Overall, most predicted structures were within an RMSD range of 2–3~\AA, with each 300~ns simulation requiring approximately 10–12~hours of computation on a multi-core workstation equipped with NVIDIA GeForce GTX~1080~Ti GPUs using GPU-enabled Gromacs-2024 version \cite{gromacs_2024}. Although extending the simulation to the microsecond scale could improve accuracy, it becomes increasingly time-consuming, especially for longer peptide chains.

The results obtained from IBM’s quantum hardware were also in close agreement with the classical MD outcomes. However, the key advantage of the quantum approach lies in its faster sampling efficiency, showing promising potential for studying shorter peptides and motivating future work toward improving its scalability and accuracy.




\section{\textbf{Conclusions} \label{conclusions}}

In this paper, we have presented a novel turn-based encoding scheme for predicting the backbone structures of protein sequences using the Variational Quantum Eigensolver implemented on quantum computing hardware. Our approach leverages a hybrid quantum-classical optimization framework and incorporates the Miyazawa–Jernigan potential to model inter-residue interactions, allowing realistic folding predictions on a three-dimensional face-centered cubic lattice. Comparison with experimental structures shows that the predicted conformations achieve RMSD values within $3.2\AA$. Our results demonstrate the potential of hybrid quantum-classical algorithms for tackling the protein folding problem, particularly for small to medium-sized peptides, and provide a foundation for future exploration of more complex proteins as quantum hardware continues to advance. 
Future work will focus on refining encoding strategies to incorporate explicit side-chain and solvent interactions, improving the accuracy of predicted structures further, and extending the approach beyond static conformations to model the pathways of protein folding.

\bibliographystyle{IEEEtran}

\begin{thebibliography}{10}
\providecommand{\url}[1]{#1}
\csname url@samestyle\endcsname
\providecommand{\newblock}{\relax}
\providecommand{\bibinfo}[2]{#2}
\providecommand{\BIBentrySTDinterwordspacing}{\spaceskip=0pt\relax}
\providecommand{\BIBentryALTinterwordstretchfactor}{4}
\providecommand{\BIBentryALTinterwordspacing}{\spaceskip=\fontdimen2\font plus
\BIBentryALTinterwordstretchfactor\fontdimen3\font minus \fontdimen4\font\relax}
\providecommand{\BIBforeignlanguage}[2]{{%
\expandafter\ifx\csname l@#1\endcsname\relax
\typeout{** WARNING: IEEEtran.bst: No hyphenation pattern has been}%
\typeout{** loaded for the language `#1'. Using the pattern for}%
\typeout{** the default language instead.}%
\else
\language=\csname l@#1\endcsname
\fi
#2}}
\providecommand{\BIBdecl}{\relax}
\BIBdecl

\bibitem{protein_human}
Z.-l. Li and M.~Buck, ``Beyond history and “on a roll”: The list of the most well-studied human protein structures and overall trends in the protein data bank,'' \emph{Protein Science}, vol.~30, no.~4, pp. 745--760, 2021.

\bibitem{UniProt}
\BIBentryALTinterwordspacing
T.~U. Consortium, ``Uniprot: the universal protein knowledgebase in 2025,'' \emph{Nucleic Acids Research}, vol.~53, no.~D1, pp. D609--D617, 11 2024. [Online]. Available: \url{https://doi.org/10.1093/nar/gkae1010}
\BIBentrySTDinterwordspacing

\bibitem{PDB}
\BIBentryALTinterwordspacing
B.~K. Stephen, D.~W. Piehl, B.~Vallat, and C.~Zardecki, ``{RCSB Protein Data Bank: supporting research and education worldwide through explorations of experimentally determined and computationally predicted atomic level 3D biostructures},'' \emph{IUCrJ}, vol.~11, no.~3, pp. 279--286, May 2024. [Online]. Available: \url{https://doi.org/10.1107/S2052252524002604}
\BIBentrySTDinterwordspacing

\bibitem{alphafold}
J.~Abramson, J.~Adler, J.~Dunger, R.~Evans, T.~Green, A.~Pritzel, O.~Ronneberger, L.~Willmore, A.~J. Ballard, J.~Bambrick, and et~al., ``Accurate structure prediction of biomolecular interactions with alphafold 3,'' \emph{Nature}, vol. 630, no. 8016, p. 493–500, May 2024.

\bibitem{rosetta}
M.~Baek, F.~DiMaio, I.~Anishchenko, J.~Dauparas, S.~Ovchinnikov, G.~R. Lee, J.~Wang, Q.~Cong, L.~N. Kinch, R.~D. Schaeffer, and et~al., ``Accurate prediction of protein structures and interactions using a three-track neural network,'' \emph{Science}, vol. 373, no. 6557, p. 871–876, Aug 2021.

\bibitem{MD}
\BIBentryALTinterwordspacing
J.~A. McCammon, B.~R. Gelin, and M.~Karplus, ``Dynamics of folded proteins,'' \emph{Nature}, vol. 267, no. 5612, pp. 585--590, Jun 1977. [Online]. Available: \url{https://doi.org/10.1038/267585a0}
\BIBentrySTDinterwordspacing

\bibitem{Levinthal}
\BIBentryALTinterwordspacing
{Levinthal, Cyrus}, ``Are there pathways for protein folding?'' \emph{J. Chim. Phys.}, vol.~65, pp. 44--45, 1968. [Online]. Available: \url{https://doi.org/10.1051/jcp/1968650044}
\BIBentrySTDinterwordspacing

\bibitem{FCC}
\BIBentryALTinterwordspacing
Z.~Bagci, R.~L. Jernigan, and I.~Bahar, ``Residue coordination in proteins conforms to the closest packing of spheres,'' \emph{Polymer}, vol.~43, no.~2, pp. 451--459, 2002. [Online]. Available: \url{https://www.sciencedirect.com/science/article/pii/S003238610100427X}
\BIBentrySTDinterwordspacing

\bibitem{MJpot}
\BIBentryALTinterwordspacing
S.~Miyazawa and R.~L. Jernigan, ``Residue – residue potentials with a favorable contact pair term and an unfavorable high packing density term, for simulation and threading,'' \emph{Journal of Molecular Biology}, vol. 256, no.~3, pp. 623--644, 1996. [Online]. Available: \url{https://www.sciencedirect.com/science/article/pii/S002228369690114X}
\BIBentrySTDinterwordspacing

\bibitem{resAnalysis_Linn}
\BIBentryALTinterwordspacing
H.~Linn, I.~Lyngfelt, L.~Garc\'{\i}a-\'Alvarez, and G.~Johansson, ``Resource analysis of quantum algorithms for coarse-grained protein folding models,'' \emph{Phys. Rev. Res.}, vol.~6, p. 033112, Jul 2024. [Online]. Available: \url{https://link.aps.org/doi/10.1103/PhysRevResearch.6.033112}
\BIBentrySTDinterwordspacing

\bibitem{Guzik_protein}
A.~Perdomo, C.~Truncik, I.~Tubert-Brohman, G.~Rose, and A.~Aspuru-Guzik, ``Construction of model hamiltonians for adiabatic quantum computation and its application to finding low-energy conformations of lattice protein models,'' \emph{Phys. Rev. A}, vol.~78, p. 012320, Jul 2008.

\bibitem{Babbush_2014}
R.~Babbush, A.~Perdomo-Ortiz, B.~O{\textquotesingle}Gorman, W.~Macready, and A.~Aspuru-Guzik, ``Construction of energy functions for lattice heteropolymer models: Efficient encodings for constraint satisfaction programming and quantum annealing,'' in \emph{Advances in Chemical Physics}.\hskip 1em plus 0.5em minus 0.4em\relax Wiley, Apr 2014, pp. 201--244.

\bibitem{Anton_2021}
A.~Robert, P.~Barkoutsos, S.~Woerner, and I.~Tavernelli, ``Resource-efficient quantum algorithm for protein folding,'' \emph{npj Quantum Information}, vol. 7:38, 2021.

\bibitem{Babej_2018}
M.~Fingerhuth, T.~Babej, and C.~Ing, ``A quantum alternating operator ansatz with hard and soft constraints for lattice protein folding,'' \emph{arXiv preprint arXiv:1810.13411}, 2018.

\bibitem{doi:10.1021/acs.jctc.4c00848}
\BIBentryALTinterwordspacing
J.~V. Pamidimukkala, S.~Bopardikar, A.~Dakshinamoorthy, A.~Kannan, K.~Dasgupta, and S.~Senapati, ``Protein structure prediction with high degrees of freedom in a gate-based quantum computer,'' \emph{Journal of Chemical Theory and Computation}, vol.~20, no.~22, pp. 10\,223--10\,234, 2024, pMID: 39504453. [Online]. Available: \url{https://doi.org/10.1021/acs.jctc.4c00848}
\BIBentrySTDinterwordspacing

\bibitem{dwaveQubo_Irback}
\BIBentryALTinterwordspacing
A.~Irb\"ack, L.~Knuthson, and S.~Mohanty, ``Folding lattice proteins confined on minimal grids using a quantum-inspired encoding,'' \emph{Phys. Rev. E}, vol. 112, p. 045302, Oct 2025. [Online]. Available: \url{https://link.aps.org/doi/10.1103/8n7p-7lh2}
\BIBentrySTDinterwordspacing

\bibitem{Omar_paper_FCC}
e.~a. Li, Rui-Hao, ``Quantum algorithm for protein structure prediction using the face-centered cubic lattice,'' \emph{arXiv preprint arXiv:2507.08955}, 2025.

\bibitem{fccEncoding}
\BIBentryALTinterwordspacing
K.~Dasgupta, ``Encoding lattice structures in quantum computational basis states,'' \emph{arXiv:2406.01547v1}, 2024. [Online]. Available: \url{https://doi.org/10.48550/arXiv.2406.01547}
\BIBentrySTDinterwordspacing

\bibitem{Barkoutsos-cVar}
\BIBentryALTinterwordspacing
P.~K. Barkoutsos, G.~Nannicini, A.~Robert, I.~Tavernelli, and S.~Woerner, ``Improving variational quantum optimization using {CVaR},'' \emph{Quantum}, vol.~4, p. 256, April 2020. [Online]. Available: \url{https://doi.org/10.22331%2Fq-2020-04-20-256}
\BIBentrySTDinterwordspacing

\bibitem{dist_cutoff}
R.~Nagarajan, A.~Archana, A.~M. Thangakani, S.~Jemimah, D.~Velmurugan, and M.~M. Gromiha, ``Pdbparam: Online resource for computing structural parameters of proteins,'' \emph{Bioinform Biol Insights. 2016 Jun 14;10:73-80}, 2016.

\bibitem{chimera}
E.~F. Pettersen, T.~D. Goddard, C.~C. Huang, G.~S. Couch, D.~M. Greenblatt, E.~C. Meng, and T.~E. Ferrin, ``Ucsf chimera—a visualization system for exploratory research and analysis,'' \emph{Journal of computational chemistry}, vol.~25, no.~13, pp. 1605--1612, 2004.

\bibitem{v_rescale}
G.~Bussi, D.~Donadio, and M.~Parrinello, ``Canonical sampling through velocity rescaling,'' \emph{The Journal of chemical physics}, vol. 126, no.~1, 2007.

\bibitem{parrinello1981polymorphic}
M.~Parrinello and A.~Rahman, ``Polymorphic transitions in single crystals: A new molecular dynamics method,'' \emph{Journal of Applied physics}, vol.~52, no.~12, pp. 7182--7190, 1981.

\bibitem{amber_ff}
K.~Lindorff‐Larsen, S.~Piana, K.~Palmo, P.~Maragakis, J.~L. Klepeis, R.~O. Dror, and D.~E. Shaw, ``Improved side‐chain torsion potentials for the amber ff99sb protein force field,'' \emph{Proteins: Structure, Function, and Bioinformatics}, vol.~78, no.~8, p. 1950–1958, Apr 2010.

\bibitem{gromacs_2024}
\BIBentryALTinterwordspacing
M.~Abraham, A.~Alekseenko, V.~Basov, C.~Bergh, E.~Briand, A.~Brown, M.~Doijade, G.~Fiorin, S.~Fleischmann, S.~Gorelov, G.~Gouaillardet, A.~Grey, M.~E. Irrgang, F.~Jalalypour, J.~Jordan, C.~Kutzner, J.~A. Lemkul, M.~Lundborg, P.~Merz, V.~Miletic, D.~Morozov, J.~Nabet, S.~Pall, A.~Pasquadibisceglie, M.~Pellegrino, H.~Santuz, R.~Schulz, T.~Shugaeva, A.~Shvetsov, A.~Villa, S.~Wingbermuehle, B.~Hess, and E.~Lindahl, ``Gromacs 2024.3 manual,'' Aug. 2024. [Online]. Available: \url{https://doi.org/10.5281/zenodo.13457083}
\BIBentrySTDinterwordspacing

\bibitem{qiskit}
A.~Javadi-Abhari, M.~Treinish, K.~Krsulich, C.~J. Wood, J.~Lishman, J.~Gacon, S.~Martiel, P.~D. Nation, L.~S. Bishop, A.~W. Cross, B.~R. Johnson, and J.~M. Gambetta, ``Quantum computing with qiskit,'' \emph{arXiv preprint arxiv.org/abs/2405.08810}, 2024.

\bibitem{mthree}
P.~D. Nation, H.~Kang, N.~Sundaresan, and J.~M. Gambetta, ``Scalable mitigation of measurement errors on quantum computers,'' \emph{PRX Quantum 2 040326}, 2021.

\end{thebibliography}

\end{document}